\documentclass[aps,prb,superscriptaddress,twocolumn,floatfix]{revtex4}
\usepackage{color,graphicx,pstricks}
\usepackage{psfrag}
\usepackage{textcomp}
\usepackage{hyperref}
\hypersetup{
  colorlinks=true,
  citecolor=blue,
  linkcolor=red,
  }
\usepackage{amsmath}
\begin{document}

\title {Microscopic study of interlayer magnetic coupling across the interface in 
  antiferromagnetic bilayers}

\author{Sandip Halder}
\affiliation{Theory Division, Saha Institute of Nuclear Physics,
  A CI of Homi Bhabha National Institute, Kolkata-700064, India}
\author{Sudip Mandal}
\affiliation{Theory Division, Saha Institute of Nuclear Physics,
  A CI of Homi Bhabha National Institute, Kolkata-700064, India}
\author{Kalpataru Pradhan}
\email{kalpataru.pradhan@saha.ac.in}
\affiliation{Theory Division, Saha Institute of Nuclear Physics,
  A CI of Homi Bhabha National Institute, Kolkata-700064, India}

\begin{abstract}
The enhancement of Neel temperature ($T_N$) of low-$T_{N}$ antiferromagnets in
antiferromagnetic bilayers AF1/AF2, where the $T_N$ of AF1 is larger than AF2 (for example
enhancement of $T_{N}$ of CoO in CoO/NiO or FeO in FeO/CoO), is a subject of considerable
interest. One essential question needs to be answered in these bilayers: is the interfacial
coupling a short-range one or long-range that mediates the effect of the AF1 layers on the
magnetic properties of AF2 layer? To understand the systematics of the magnetic coupling
across the interface, we investigate the plane-resolved magnetotransport properties of
antiferromagnetic bilayers using an electron-hole symmetric one-band Hubbard model at
half-filling, employing a semi-classical Monte Carlo method. In our model Hamiltonian
calculations, we set Coulomb repulsion $U_{1} = 8$ to mimic high-$T_{N}$ AF1 layer, whereas
we use $U_{2}$ = $2\times U_{1}$ to mimic the low-$T_{N}$ AF2 layer. Our calculations
show that the $T_{N}$ of the low-$T_{N}$ antiferromagnet enhances substantially when
it's thickness is small, similar to experiments, giving rise to single magnetic transition
temperature of the bilayer system. These findings are well supported by a single peak
in temperature-dependent specific heat. However, for larger thicknesses, the $T_{N}$
of the low-$T_{N}$ antiferromagnet approaches towards its bulk value and constituent
antiferromagnetic layers align antiferromagnetically at two separate temperatures and
two maxima are observed in specific heat data. Our calculations also show that the
delocalization of moments is more or less confined near the interface indicating the
short-ranged nature of the proximity effect. Our obtained results are consistent with
the experimental observations. A detailed discussion of the modifications that will
occur if we use $U_{1} = 8$ and $U_{2}$ = $0.5\times U_{1}$ will also be addressed.
\end{abstract}

\maketitle

\section{Introduction}
Antiferromagnets (AFs) based on transition metal oxides (TMOs) have garnered renewed
interest for their potential in future spintronic devices due to their compelling
physical properties~\cite{Fina,XMarti,Wadley1,Wadley2}. Their insensitivity to  magnetic
fields makes AFs promising candidate for designing magnetic data storage~\cite{Mangin,Sbiaa,Ikeba,Nemec,Song}
and magneto-electronic devices~\cite{Wolf}, as they do not create stray fields, thereby
enhancing the scalability of the technology of magnetic memory devices~\cite{Baltz,Hou}.
In fact, zero magnetic moment of AFs makes them suitable for designing high-density
memory integration in spintronic devices~\cite{Park,Wang,XMarti}. Subsequently, these
AF materials are also useful in advanced random-access memories, magnetic recording and
sensor devices~\cite{Nogues}. Overall, the physics of AF materials is extremely rich,
and occasionally distinct and surprising from their ferromagnetic counterparts.

Despite of the rich physics and numerous practical applications, the majority of antiferromagnetic
materials now appear to be inappropriate for use because of their low transition temperature.
In order to get around the restriction of operating temperatures in devices, raising the
antiferromagnet's transition temperature $T_N$ has thus emerged as a ongoing area of
study~\cite{Kaziol,Baldrati}. The $T_{N}$ of AF can be enhanced by implementing a number of
techniques, such as: doping, strain, and  heterostructure engineering. For instance, 
$T_{N}$ of the magnetoelectric antiferromagnet $Cr_{2}O_{3}$ is enhanced by 10\% with just 1\%
substitution of oxygen by boron anion~\cite{SMu,Street}. In $Cr_{2}O_{3}$, the antiferromagnetic
order is established by Cr-Cr direct exchange coupling~\cite{Shi}. Boron doping introduces impurity
states that mediate additional strong hibridization between the neighboring Cr ions. As a result, 
the $T_{N}$ of the $Cr_{2}O_{3}$ increases, making it suitable for room temperature 
applications~\cite{XHe}. Furthermore, the $T_{N}$ of the AF $YMnO_{3}$ can be enhanced by
controlling the strain, achieved by growing $YMnO_{3}$ on different substrates~\cite{KHWu}.
Additionally, by designing ultra-thin heterostructures such as $[(LaMnO_{3})_{n}/(SrMnO_{3})_{2n}]_{m}$ 
(with $n = 1$ and $2$), the Neel temperature of the resulting C-type antiferromagnet is increased 
by $~\sim 70 K$ as compared to the solid solution $La_{1/3}Sr_{2/3}MnO_{3}$~\cite{May}.

Similar to the enhancement of the ferromagnetic Curie temperature $T_{C}$ of diluted
magnetic semiconductors when they come into contact with FMs~\cite{Maccherozzi,Olejnik,Manna}
or AFs~\cite{Eid,Manna}, the proximity effect can also be used as an efficient way to increase
the $T_{N}$ of AF materials. It has also been found that the magnetic proximity effect
greatly increases the ferromagnetic $T_C$ of FM in FM/FM bilayers~\cite{Choi,Vargas,JWu}.
Additionally, embedding a ferromagnetic Co nanocluster inside a high $T_{C}$ ferromagnetic
NiFe matrix also increases its $T_{C}$~\cite{Hindmarch}. The Co nanoclusters' magnetic
moment was stabilized until room temperature due to the strong exchange interaction
at their interfaces with NiFe nanoclusters.

The magnetic vicinity of antiferromagnetic or ferromagnetic (FM) materials actually
raises the transition temperature of AF materials, as shown by investigations based
on AF/AF~\cite{Jena,Jena1,Li,JZhu} or FM/AF~\cite{Lierop,Lenz} bilayers. So, the
augmentation of Neel temperature of one of the AF material in AF/AF bilayers, particularly
constructed with correlated mono-oxide materials and magnetically connected via their
interfaces, is a subject of considerable interest. In particular, the bilayer comprising $NiO$
($T_{N} = 523\;K$~\cite{Nagamiya}, High-$T_{N}$ AF) and CoO ($T_{N} = 293\; K$~\cite{Sievers},
low-$T_{N}$ AF) is the most well-known system that has been investigated~\cite{Manna} extensively
in the last few decades. For smaller thickness of the CoO layer (less than 20 \AA), the NiO/CoO
bilayer acts as a single-phase magnetic system with a single $T_{N}$~\cite{Jena1,JZhu,Borchers1,Takano}.
These experimental results show that the $T_{N}$ of the CoO layer increases considerably for smaller
thickness of the CoO layer and approaches to the bulk $T_{N}$ of NiO. But, the $T_{N}$ decreases
with increase of CoO layer thickness, and as a result, both the NiO and CoO show AF transitions
at different temperatures. Ultimately, the $T_{N}$ of CoO layer approaches to the bulk limit
($T_{N} = 293 K$) for large thickness. So, overall the $T_{N}$ of the bilayer system can be
tuned between the temperatures of $T_{N}$ of bulk NiO and CoO. A similar phenomenon of the
enhancement of Neel temperature is also evidenced in bilayers composed of rutile-type
antiferromagnets, such as in $FeF_{2}$/$CoF_{2}$ bilayer~\cite{Ramos,Carrigo}.

Direct measurements of specific heat of NiO/CoO superlattices~\cite{Abarra1,Abarra2} show a single
peak similar to $Ni_{0.5}Co_{0.5}O$ alloy for smaller thicknesses that corresponds well with the
single magnetic transition temperature of the bilayer system. For thicker bilayers, two broad maxima
are observed which are due to different Neel temperatures of the individual NiO and CoO layers~\cite{Abarra2}.
As the thickness of the bilayer increases, these two broad maxima eventually approach the individual
bulk $T_N$ of CoO and NiO. Several reports on NiO/CoO bilayers~\cite{Carey1,Carey2,Fujikata,Abarra2,Borchers1} 
establish an existence of magnetic couplings at the interface of NiO/CoO superlattices well above
the transition temperature $T_N$ of CoO.

As previously stated, the enhancement of low-$T_N$ antiferromagnets is not limited to
antiferromagnetic bilayers.
The Neel temperature of the antiferromagnet ($Co_{3}O_{4},\; T_{N} \sim 40\;K$~\cite{Roth}) also gets
enhanced due to the proximity effect of the higher $T_{C}$ ferromagnet ($Ni_{80}Fe_{20},\; T_{N} \sim 800\;K$~\cite{Handley})
in $Ni_{80}Fe_{20}$/$Co_{3}O_{4}$ bilayer~\cite{Lierop}.
The Neel temperature of CoO and NiO also enhances in $Fe_{3}O_{4}/CoO$~\cite{Zaag}
and $Fe_{3}O_{4}/NiO$~\cite{Borchers2} thin film superlattices, where $Fe_{3}O_{4}$ 
is a ferrimagnet with high ferrimagnetic curie temperature $T_{c}\sim 858\;K$. This
enhancement of Neel temperature takes place in the bilayer structure, is
associated with the magnetic coupling due to the proximity effect in contact with
high transition temperature $Fe_{3}O_{4}$. Magnetization data of the individual
planes of CoO (not in contact with $Fe_{3}O_{4}$, i.e., planes which resides far away
from the interfaces) shows no enhancement of $T_{N}$, signifies the effect of $Fe_{3}O_{4}$ on
the Neel temperature enhancement through the proximity effect is confined to very limited
region around the interfaces. There are also evidences 
of increase orbital magnetic moment of Ni and Fe in the interfaces of $Fe_{3}O_{4}/NiO$ 
superlattice~\cite{Krug}. An exchange interaction is also reported at the interfaces~\cite{Zaag,Ball,Gatel}
of antiferromagnet (CoO or NiO) and ferrimagnet ($Fe_{3}O_{4}$) superlattices, which shifts
the magnetic hysteresis loop along (or, opposite) to the magnetic field axis, resulted in an
exchange bias phenomena. The observed exchange bias field ($H_{EB}$) decreases with increase
of temperature and vanishes just above the exchange bias blocking temperature.
On the other hand, in the core-shell MnO ($T_{N}\sim 120\;K$)/$
\gamma$-$Mn_{2}O_{3}$ ($T_{C}\sim 40\; K$) nanocomposites, the magnetic moment in the ferrimagnetic
shell $\gamma$-$Mn_{2}O_{3}$ is stable far above $T_{C}$ (up to the $T_{N}$ of MnO) due to the magnetic
proximity effect~\cite{Golosovsky}, which is interesting for it's unconventional nature.
Moreover, in core MnO - shell $Mn_{3}O_{4}$ nanoparticles, exchange bias effect is evidenced above 
the ferrimagnetic $T_{C}$ of $Mn_{3}O_{4}$~\cite{Berkowitz}.

What are the benefits of improving the $T_{N}$ of the low $T_N$ AF layer in AF/AF
bilayers? For instance, consider the wustite antiferromagnet FeO ($T_{N} \sim 198\;K$~\cite{Koike}).
Fe/FeO bilayers exhibit an exchange bias, shifting the hysteresis loop opposite to the
field axis due to exchange interactions at the interface below the $T_{N}$ of bulk FeO~\cite{Kaziol1,Kaziol}.
However, because of its low $T_{N}$, its potential use in devices
is constrained despite its fascinating phenomenon. Growing FeO on CoO in FeO/CoO
bilayers enhances its $T_{N}$ of the order of $\sim 100\;K$ due to proximity interaction
with CoO~\cite{Kaziol}. So, the close association of CoO and FeO can overcome temperature
constraints in device applications. In fact, CoO proximity also improves the exchange interaction
between Fe and FeO in Fe/FeO bilayers, resulting in higher exchange bias and coercive field in the
Fe/FeO/CoO heterostructure. Additionally, the thickness of the CoO layer significantly
affects the exchange bias field ($H_{EB}$) and blocking temperature ($T_{B}$)~\cite{Kaziol,Malozemoff,Nowak,Binek}.
Overall, the proximity of CoO to FeO helps overcome temperature constraints in device applications.

Due to their technological significance and the prospect of expanding fundamental understanding
the physics of (low-$T_{N}$-AF)/(high-$T_{N}$-AF) bilayers, particularly at the interface, is
of tremendous interest. A microscopic knowledge is required for further advancement
in this area. Microscopic analysis can help to determine the number of atomic planes involved in
coupling across an interface or if both systems are fully engaged. So, in this article, we aim
to understand how the low-$T_{N}$ antiferromagnetic layer's Neel temperature increases
in low-$T_{N}$-AF/high-$T_{N}$-AF bilayer systems, focusing on the effect of local moment
localization/delocalization at the interface. We have employed one-band Hubbard model to
investigate the bilayers using semi-classical Monte Carlo (s-MC) technique. We find that the
$T_{N}$ of the low-$T_{N}$ antiferromagnet is enhanced significantly for smaller thickness
of the low-$T_{N}$ antiferromagnet. Subsequently, the $T_{N}$ of the low-$T_{N}$ antiferromagnet
decreases with increasing it’s thicknesses and ultimately approaches to the corresponding bulk
limit for larger thicknesses of the low-$T_{N}$ antiferromagnet. For smaller thicknesses of
the low-$T_{N}$ antiferromagnet, the moment delocalization affects its nature, and as a result,
the high-$T_{N}$ antiferromagnet strongly influences the low-$T_{N}$ antiferromagnet in
increasing the Neel temperature.

The article is arranged as follows: In Sec. \textbf{II}, we describe our reference model
Hamiltonian and numerical approach for analyzing the properties of antiferromagnetic bilayers.
We provide a quick overview of the various physical parameters in Sec. \textbf{III} that
will be utilized to investigate the magnetic and transport properties of various bilayers.
The set of parameter values to build the AF1/AF2 bilayers is subsequently identified in
Sec. \textbf{IV}. We describe the magnetotransport properties of AF1/AF2 bilayers
in Secs. \textbf{V} and \textbf{VI} in order to shed light on the phenomenon of enhancing
the Neel temperature of low-$T_N$ antiferromagnets. Our findings are summarized in
Sec. \textbf{VII}.

\section{Reference model and Methodology}
In order to investigate the magnetotransport properties of the (low-$T_{N}$-AF)/(high-$T_{N}$-AF)
bilayers, we consider following electron-hole symmetric one-band Hubbard Hamiltonian:
\begin{align}
H=&-t\sum_{<i,j>,\sigma}(c_{i,\sigma}^{\dagger}c_{j,\sigma}+H.c.)\nonumber\\
&+U\sum_{i}(n_{i,\uparrow}-\frac{1}{2})(n_{i,\downarrow}-\frac{1}{2})-\mu\sum_{i}n_{i},\nonumber 
\end{align}

where $c_{i,\sigma}^{\dagger}\;(c_{i,\sigma})$ denote the electron creation
(annihilation) operator at site $i$ with spin $\sigma\;(\uparrow or,
\downarrow)$. $t$ is the hopping amplitude between the nearest neighbors
sites. $U$ is the strength of on-site Coulomb repulsion at site $i$.
$n_{i}=c_{i,\sigma}^{\dagger}c_{i,\sigma}$ represents the occupation number
operator at site $i$. $\mu$ is the chemical potential which controls the
overall density of the system. At half-filling ($n = 1$), $\mu = 0$ in our 
electron-hole symmetric model Hamiltonian.

Next, eliminating the constant term, we express the Hamiltonian in the following 
way 
\begin{align}
H=&-t\sum_{<i,j>,\sigma}(c_{i,\sigma}^{\dagger}c_{j,\sigma}+H.c.)
+U\sum_{i}n_{i,\uparrow}n_{i,\downarrow}-\frac{U}{2}\sum_{i}n_{i},\nonumber \\
&=H_{0}+H_{int}\nonumber
\end{align}

where $H_{0}$ contains the non-interacting one-body quadratic part and $H_{int}$ consists of
the interacting quartic part of the model Hamiltonian. Then, to solve the Hamiltonian, we
decompose the quartic interaction term into two different quadratic terms as follows:

\begin{center}
$Un_{i,\uparrow}n_{i,\downarrow}=U\left[\frac{1}{4}n_{i}^{2}
-({\bf{S}}_{i}.{\hat{\Omega}}_{i})^{2})\right]$\\
\end{center}

where ${\bf{S}}_{i}$ is the spin vector at the $ith$ site, defined as
${\bf{S}}_{i}=\frac{\hbar}{2}\sum_{\alpha,\beta}c_{i,\alpha}^{\dagger}\sigma_{\alpha,\beta}c_{i,\beta}$
with $\hbar=1$,  and $\sigma$ is the Pauli matrices. ${\hat\Omega}_{i}$ is the arbitrary unit vector
at site $i$. Here, the decoupling is rotationally invariant. Then, the partition function of the model
Hamiltonian, $H=H_{0}+H_{int}$, is written as $Z=Tre^{-\beta H}$, where $\beta=\frac{1}{T}$ is the 
inverse temperature with Boltzmann constant $K_{B}=1$. The window $[0,\beta]$ is divided into $M$ number 
of equally spaced slices of girth $\Delta\tau$ ($\beta=M\Delta\tau$). To evaluate the partition function, 
we express  $e^{-\beta(H_{0}+H_{int})}=(e^{-\Delta\tau H_{0}}e^{-\Delta\tau H_{int}})^{M}$
up to first order in $\Delta\tau$ using Suzuki-Trotter transformation in the limit $\Delta\tau\rightarrow0$
(for very large $M$). Next, by implementing Hubbard-Stratonovich (H-S) transformation, the interacting 
part of the partition function $e^{-\Delta\tau U\sum_{i}[\frac{1}{4}n_{i}^{2}-({\bf{S}}_{i}.{\hat{\Omega}}_{i})^{2}]}$ 
can be shown to be proportional to 
\begin{align}
\sim\int &d\phi_{i}(l) d\Delta_{i}(l)d^{2}\Omega_{i}(l) \nonumber\\
&\times e^{-\Delta\tau [\sum_{i}\{\frac{{\phi_{i}(l)}^{2}}{U}
+i\phi_{i}(l)n_{i}+\frac{{\Delta_{i}(l)}^{2}}{U}
-i\Delta_{i}(l){\hat{\Omega}}_{i}(l).{\bf{S}}_{i}\}],}\nonumber
\end{align}

for a generic time slice `$l$'. Here, the H-S auxiliary fields $\phi_{i}(l)$ is
coupled with charge density $n_{i}$, and  $\Delta_{i}(l)$ is coupled with the 
spin vector ${\bf{S}_{i}}$. Introducing a new vector auxiliary field  
${\bf{m}}_{i}(l)=\Delta_{i}(l).{\hat\Omega}_{i}(l)$, we evaluate the total partition 
function as
\begin{align}
Z=const. & \times  Tr \prod_{l=M}^{1} \int d\phi_{i}(l)d^{3}{\bf{m}}_{i}(l)\times \nonumber\\
& e^{-\Delta\tau[H_{0}+\sum_{i}\{\frac{{\phi_{i}(l)}^{2}}{U}
+i\phi_{i}(l)n_{i}+\frac{{{\bf{m}}_{i}(l)}^{2}}{U}
-2{\bf{m}}_{i}.{\bf{S}}_{i}\}]}\nonumber
\end{align}
where the product follows the time order product. $l$ runs from $M$ to $1$. From the partition function,
we extract an effective model Hamiltonian. At this moment, we eliminate
the $\tau$ dependence of the classical auxiliary fields and retaining the 
spatial fluctuations of the auxiliary fields. Then, we use the saddle point 
value of the auxiliary field $i\phi_{i}(l)=\frac{U}{2}<n_{i}>$. After redefining
${\bf{m}}_{i}\rightarrow\frac{U}{2}{\bf{m}}_{i}$, we write down the effective
Hamiltonian~\cite{Mukherjee,Chakraborty1,Halder1} as follows

\begin{align}
H_{eff}=& -t\sum_{<i,j>,\sigma}(c_{i,\sigma}^{\dagger}c_{j,\sigma}+H.c.)\nonumber\\
&+\frac{U}{2}\sum_{i}(<n_{i}>n_{i}-{\bf{m}}_{i}.\sigma_{i})\nonumber \\
& +\frac{U}{4}\sum_{i}({{\bf{m}}_{i}}^{2}-{<n_{i}>}^{2})-\frac{U}{2}\sum_{i}n_{i}-\mu\sum_{i}n_{i}.
\end{align}

We deal with this spin-fermionic effective model Hamiltonian by diagonalizing 
the fermionic sector in the fixed background of classical auxiliary fields
$\{{\bf{m}}_{i}\}$ and $\{<n_{i}>\}$. During the Monte Carlo (MC) update, we visit
every lattice site sequentially and sampling the classical auxiliary fields $\{{\bf{m}}_{i}\}$
using standard Metropolis algorithm. We evaluate $\{n_{i}\}$ self-consistently 
at every $10th$ step of the MC system sweep. We use $2000$ MC system sweeps at a fixed
temperature, where the
first $1000$ MC sweeps  are used to thermalize the system and the remaining $1000$ MC 
sweeps are devoted to measuring the physical observables. We discard $10$ MC sweeps
between the measurements to avoid illicit self-correlation in the data. We access
large system sizes by implementing travelling cluster approximation based Monte Carlo 
technique~\cite{Kumar,Pradhan,Pradhan1,Chakraborty,Halder}.

\section{Physical Observables}

In order to study the magnetotransport properties of the high-$T_N$-AF1/low-$T_N$-AF2 bilayers,
we measure various physical observables. Specifically, we calculate the following structure factor
associated with quantum spin correlations to estimate the Neel temperatures:
\begin{align}
S(\boldsymbol{q})=\frac{1}{N^{2}}\sum_{i,j}\langle {\boldsymbol{S}}_{i}.{\boldsymbol{S}}_{j} \rangle
e^{-i{\boldsymbol{q}}.({\boldsymbol{r}}_{i}-{\boldsymbol{r}}_{j})}, \nonumber
\end{align}
where $\bf{q}$ is the wave vector and $N$ is the total number of lattice sites in the system.
$\mathbf{S_{i}}$ is the quantum spin at site $i$, calculated using the eigen values and eigen vectors 
of the effective Hamiltonian. $i$ and $j$ run all over the lattice sites. The angular brackets denote
the thermal and quantum mechanical averages of the observables over the Monte Carlo generated
equilibrium configurations, along with the configurational averages over the ten initial
configurations of the classical auxiliary fields. 

The specific heat of the bilayers is calculated by differentiating the total energy of the 
system with respect to temperature, $C_{v}(U,T) = \frac{dE(U,T)}{dT}$. The central difference formula
is applied to estimate the specific heat numerically. We also evaluate the average local moment
of the system (a measure of the system averaged magnetization squared)
by using the formula: 
$M = \langle (n_{\uparrow}-n_{\downarrow})^{2}\rangle = \langle n \rangle -2\langle n_{\uparrow}n_{\downarrow}\rangle$, 
where $\langle n \rangle = \langle n_{\uparrow}+n_{\downarrow}\rangle$.

We estimate the density of states (DOS) at a particular frequency $\omega$, which is
defined as $DOS(\omega) = \frac{1}{N}\sum_{\alpha}\delta (\omega - \epsilon_{\alpha})$, where 
$\epsilon_{\alpha}$ is the single particle eigen values, and $\alpha$ runs over the total
number ($= 2N$) of eigen values of the system. We implement a Lorentzian representation of the
delta function with broadening $\sim BW/2N$ (where $BW$ is the bare bandwidth) to enumerate
the DOS.

In addition, we calculate the out-of-plane (along z-axis) and in-plane (along x-axis) conductivities 
of the AF1/AF2 bilayers in the dc limit using the Kubo-Greenwood formalism~\cite{Mahan,Kumar1,Bulanchuk},
which is represented by

\[\sigma(\omega)=\frac{A}{N}\sum_{\alpha, \beta}(n_{\alpha}-n_{\beta})
\frac{{|f_{\alpha \beta}|}^{2}}{\epsilon_{\beta}-\epsilon_{\alpha}}
\delta[\omega-(\epsilon_{\beta}-\epsilon_{\alpha})]\]

\noindent
where $A=\pi e^{2}/\hbar a$ ($a$ is the lattice parameter). $f_{\alpha \beta}$
represents the matrix elements of the paramagnetic current operator
${\hat{j}}_{z}=it\sum_{i,\sigma}
(c_{i,\sigma}^{\dagger}c_{i+z,\sigma}-c_{i+z,\sigma}^{\dagger}c_{i,\sigma})$
or, ${\hat{j}}_{x}=it\sum_{i,\sigma}(c_{i,\sigma}^{\dagger}c_{i+x,\sigma}-
c_{i+x,\sigma}^{\dagger}c_{i,\sigma})$
between the eigen states $|\psi_{\alpha}>$ and $|\psi_{\beta}>$ with corresponding
eigen energies $\epsilon_{\alpha}$ and $\epsilon_{\beta}$, respectively,
and $n_{\alpha} = \theta(\mu - \epsilon_{\alpha})$ is the Fermi function
associated with the single particle energy level $\epsilon_{\alpha}$.
Next, the averaged dc conductivity, averaged over a small low-frequency interval
($\Delta \omega$), is determined as follows:
\[ \sigma_{av}(\Delta\omega)=\frac{1}{\Delta\omega}\int_{0}^{\Delta\omega}\sigma(\omega) d\omega\]
where $\Delta\omega$ is chosen three to five times larger than the mean finite size gap (average eigen value 
separation) of the system, which is actually the ratio of the bare bandwidth to the total number
of eigen values. All the physical parameters, such as $U$, $T$, and $\omega$, are measured
in units of $t$.

To understand the delocalization of local moments across the interface
of AF1/AF2 bilayers, we also evaluate the effective hopping parameter (a measure
of the gain in kinetic energy)~\cite{White,Mondaini} along the out-of-plane direction (z-axis)
as follows:
\[t_{eff}\equiv \left(\frac{t^{bilayer}}{t}\right)_{z}
=\frac{{\Big\langle\sum_{i,\sigma}(c_{i+z,\sigma}^{\dagger}c_{i,\sigma}
+c_{i,\sigma}^{\dagger}c_{i+z,\sigma})\Big\rangle}_{bilayer}}{{\Big\langle\sum_{i,\sigma}(c_{i+z,\sigma}^{\dagger}c_{i,\sigma}
+c_{i,\sigma}^{\dagger}c_{i+z,\sigma})\Big\rangle}_{0}},\]
where angular brackets represent the expectation value in the bilayer system.

\section{Parameter values to setup antiferromagnetic bilayers} 

\begin{figure}[!t]
\centerline{
\includegraphics[width=8.5cm,height=6.30cm,clip=true]{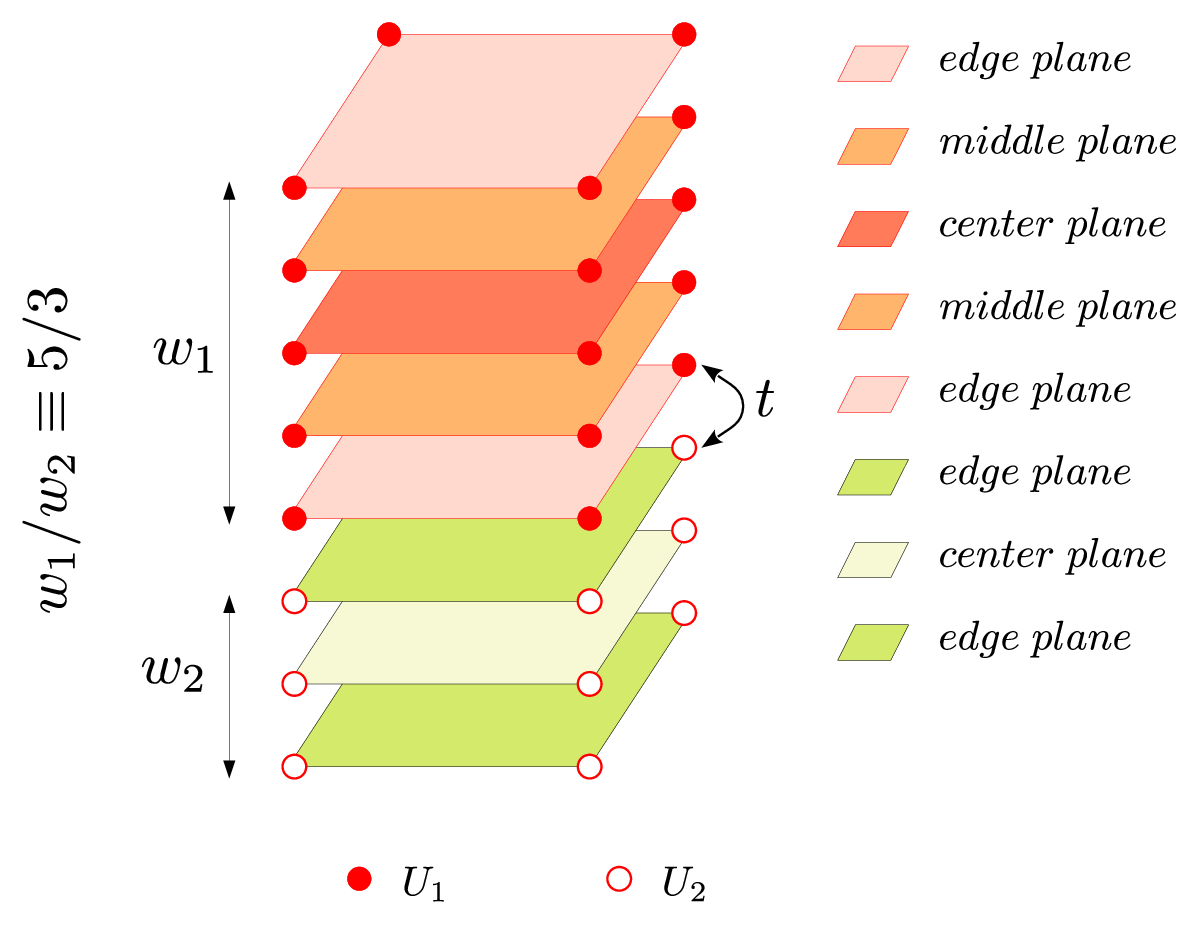}}
\caption{
Schematic representation of AF1/AF2 bilayer system. The red filled (opened) symbol indicates
the antiferromagnetic AF1 (AF2) layer. On-site Coulomb repulsion $U_{1}$ ($U_{2}$)
is assigned to mimic high-$T_N$-AF1 (low-$T_N$-AF2) layer. The antiferromagnetic AF1 (AF2)
layer has a thickness of $w_{1}$ ($w_{2}$), indicating the number of planes involved.
The bilayer system is referred to as $w_{1}$/$w_{2}$ bilayer. Different colors are used
to denote the edge, middle, and center planes of the AF1 layer. The AF2 layer's edge and
center planes are also depicted with two different colors. The hopping parameter $t$
connects the AF1 and AF2 layers at the interface. This schematic specifically depicts
a 5/3 bilayer.
} 
\label{fig01}
\end{figure}

A bilayer structure made up of two distinct antiferromagnetic layers is depicted in
Fig.~\ref{fig01}. We characterize the AF1 (AF2) layer by assigning the on-site Hubbard
repulsive strength $U_{1}$ ($U_{2}$) and thickness $w_{1}$ ($w_{2}$). We set same
hopping parameter (t) for both layers. We choose $U_{1}$ and $U_{2}$ to ensure
distinct Neel temperatures for the two layers. Essentially, in our model Hamiltonian
calculations, the AF1 (AF2) layer with thickness $w_{1}$ ($w_{2}$) is composed of $w_{1}$
($w_{2}$) 2D planes with high-$T_N$ (low-$T_N$) value. We refer to this AF1/AF2 bilayer
structure as the $w_{1}$/$w_{2}$ antiferromagnetic bilayer (or simply $w_{1}$/$w_{2}$ bilayer).
In the 5/3 (AF1/AF2) bilayer (see Fig.~\ref{fig01}), the AF1 (AF2) layer has a thickness
of $w_{1}$ = 5 and $w_{2}$ = 3. So, in the illustration of 5/3 bilayer, the AF1 layer consists
of two edge planes, two middle planes, and one center plane. At the same time the AF2 layer
contains two edge planes and one center plane. The hopping parameter (t) connects the
two layers at the interface. Periodic boundary conditions are considered in all three
directions. For 5/3 bilayer we use $8\times8\times8$ system for our calculations. In general
we set 	$8\times8\times w_T$ system, where $w_T = w_1 + w_2$ for our studies.

\begin{figure}[!t]
\centerline{
\includegraphics[width=8.5cm,height=6.30cm,clip=true]{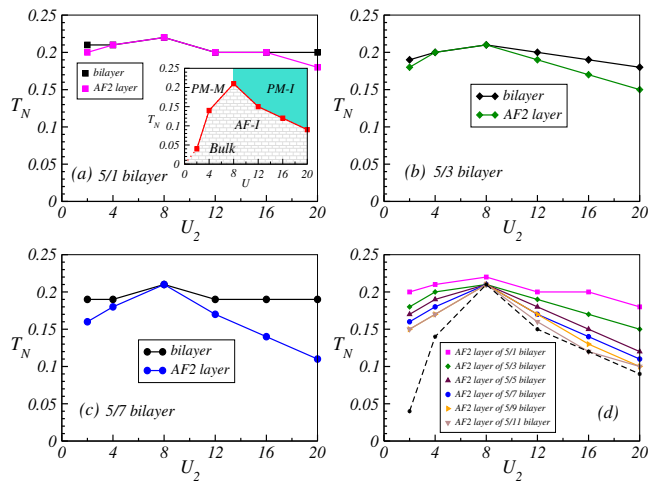}}
\caption{
The variation of Neel temperature $T_{N}$ [obtained from $S(\pi, \pi, \pi)$ vs $T$ calculations]
of the whole bilayer and the constituent AF2 layer with increasing the $U_{2}$ value in the AF2
layer is plotted for (a) 5/1 bilayer, (b) 5/3 bilayer, and (c) 5/7 bilayer. We set $U_{2} = 8$
to mimic the high-$T_N$ AF1 layer. In 5/1 bilayer, the $T_{N}$ of the AF2 layer is nearly
identical to that of the whole bilayer. This indicates that the high-$T_{N}$ AF1 layer
significantly affects the $T_{N}$ of the AF2 layer. As the thickness of the AF2 layer
increases (as one moves from 5/1 to 5/7 bilayer), the $T_{N}$ of the AF2 layer reduces
in comparison to the bilayer, specifically for higher values of $U_{2}$. The $T_{N}$ of
the AF2 layer in the 5/7 bilayer prominently exhibits nonmonotonic behavior with $U_{2}$,
similar to the bulk system. The inset of (a) displays the bulk system's $T_{N}$ vs
$U$. Here, PM-M (PM-I) represents paramagnetic metal (paramagnetic insulator). AF-I indicates 
antiferromagnetic insulating state. (d) The variation of $T_{N}$ of the AF2 layer with
$U_{2}$ for various bilayers is shown for comparison. For thicker AF2 layers
(i.e., for 5/9 and 5/11 bilayers), the AF2 layer's $T_{N}$ approaches the bulk $T_{N}$
at high $U_{2}$ values. To better visualize this characteristic, we re-plotted the
$T_{N}$ of the bulk system for different $U_{2}$ using the black dashed line.
} 
\label{fig02}
\end{figure}

First, we examine the well-studied U-T phase digram for the bulk system to qualitatively
capture the key physics of individual AF layers in bilayer systems as illustrated in
the inset of Fig.~\ref{fig02}(a). We will briefly discuss the necessary basic
characteristics of the phase diagram in order to select $U_{1}$ and $U_{2}$ values
to simulate two antiferromagnets while keeping their Neel temperatures in mind. At
low temperatures, the bulk system's ground state remains in a G-type antiferromagnetic
insulating state with finite values of $U$. The Neel temperature ($T_{N}$) exhibits
non-monotonic behavior as $U$ increases. So, the $T_{N}$ essentially rises with $U$
until $U = 8$ and at that point it reaches its optimal value $(\sim 0.21)$.
Bulk calculations are performed using $8\times8\times8$ system. After
$U = 8$, the $T_{N}$ starts to decrease as $U$ increases further. The bulk system
directly transits from a paramagnetic metallic state to an antiferromagnetic insulating
state for $U < 8$, meaning that the metal-insulator transition temperature ($T_{MIT}$)
and the Neel temperature coincide (i.e., $T_{MIT} = T_{N}$). However, for $U \ge 8$,
the change from a paramagnetic metallic state to an antiferromagnetic insulating
state occurs via a paramagnetic insulating state, i.e., $T_{MIT} > T_{N}$, as the
temperature decreases~\cite{Mukherjee,Chakraborty1}. Our findings are in good
agreement with earlier findings~\cite{Mukherjee,Chakraborty1,Rohringer,Staudt,Halder1}.

In order to replicate high-$T_{N}$ AF1 materials, we set the on-site Coulomb repulsive
strength $U_{1} = 8$ in our investigation. As mentioned above, for this value
of $U_{1}$, the bulk system's ground state stays in an antiferromagnetic insulating
state with a high $T_{N}$ ($\sim0.21$). Additionally, we fix the AF1 layer's
thickness at $w_{1} = 5$. By altering the on-site Hubbard repulsive strength
$U_{2}$ and thickness $w_{2}$ of the AF2 layer, we examine the magnetotransport
properties of bilayer systems.

We will now briefly present the modification of the bilayer's $T_N$ by altering
$U_{2}$ values. Fig.~\ref{fig02}(a) shows the $T_{N}$ of the AF2 layer of the 5/1
bilayer, as well as the $T_{N}$ of the whole bilayer system. The Neel temperatures
of the 5/1 bilayer remain steady across all $U_{2}$ values and are comparable to the
AF1 system. It's interesting to note that for all $U_{2}$ values, the $T_{N}$ of
the AF2 layer and the $T_{N}$ of the 5/1 bilayer are equal baring $U_{2}$ = 2 and
20 values. Thus, the high-$T_{N}$ AF1 layer in the 5/1 bilayer greatly raises the
$T_{N}$ of the AF2 layer.

For the 5/3 bilayer, the $T_{N}$ of the AF2 layer reduces slightly for $U_{2} > 8$,
as seen in Fig.~\ref{fig02}(b). However, the AF2 layer has a higher $T_{N}$ than
the corresponding bulk system for any specified $U_{2}$, but is lower than the bilayer's
$T_{N}$ value (unless $U_{2} = U_{1} = 8$). The $T_{N}$ of the AF2 layer exhibits
nonmonotonic behavior with $U_{2}$ for the 5/7 bilayer, as shown in Fig.~\ref{fig02}(c),
similar to bulk systems [see in set of Fig.~\ref{fig02}(a)]. Clearly, the $T_{N}$ of
the AF2 layer is less than that of the bilayer system. The $T_{N}$ of the AF2 layer in
the bilayer decreases with increasing the thicknesses, as shown in Fig.~\ref{fig02}(d).
The reduction is particularly noticeable for higher values of $U_{2}$.
The $T_{N}$ of the AF2 layer closely resembles that of the bulk system for
thicker AF2 layers (such as in 5/9 and 5/11 bilayers) and larger $U_{2}$ values,
suggesting that the proximity effect is reduced to affect the inner AF2 layers as
AF2 layer's thickness increases.

\begin{figure}[!t]
\centerline{
\includegraphics[width=8.5cm,height=6.30cm,clip=true]{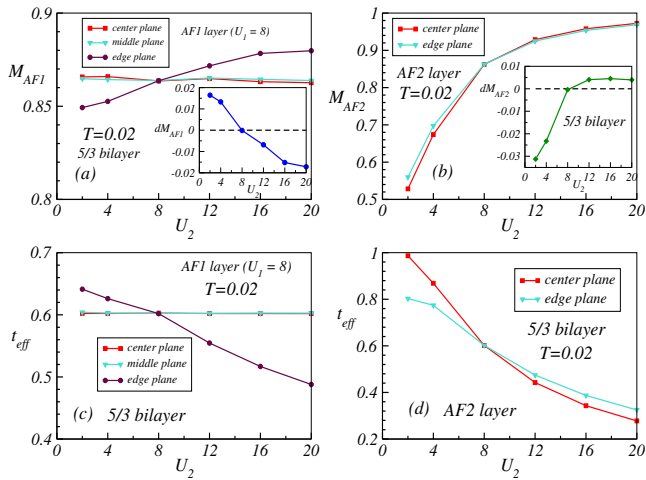}}
\caption{
(a) The local moment in the edge, middle, and center planes of the AF1 layer ($U_{1} = 8$) of
a 5/3 bilayer is plotted with varying $U_{2}$ value in the AF2 layer at $T = 0.02$.
When $U_{2} < 8$ ($U_{2} > 8$), the local moment in the edge plane of the AF1 layer is smaller
(larger) than in the center plane. The inset shows the $dM_{AF1}$, which is the difference
in average local moment between the AF1 layer's center and edge planes. $dM_{AF1}$ is positive
(negative) when $U_{2} < 8$ ($U_{2} > 8$).
(b) At $T = 0.02$, the local moments in both edge and center planes of AF2 layer
increase monotonically as $U_{2}$ increases. The local moment in the edge plane of the AF2
layer is larger (marginally smaller) than the center plane for $U_{2} < 8$ ($U_{2} > 8$).
The inset shows the $dM_{AF2}$, which represents the difference in local moment between the
AF2 layer's center and edge planes. For $U_{2} < 8$, $dM_{AF2}$ is negative, but for $U_{2} > 8$,
it's positive.
(c) The $t_{eff}$ in the edge plane of the AF1 layer decreases as the $U_{2}$ of the AF2 layer
increases, whereas the $t_{eff}$ in the middle and center planes remains nearly constant.
(d) As $U_{2}$ increases in the AF2 layer, the $t_{eff}$ in both the edge and center planes
decreases. Interestingly, the edge plane has a larger (smaller) $t_{eff}$ than the center
plane for $U_{2} < 8$ ($U_{2} > 8$) in AF1 layer whereas $t_{eff}$ of the edge plane
is smaller (larger) in AF2 layer for $U_{2} < 8$ ($U_{2} > 8$) due to interfacial coupling. 
} 
\label{fig03}
\end{figure}

To further investigate the proximity effect, we plot the magnetic moment profiles
of the 5/3 bilayer system as shown in Fig.~\ref{fig03}(a). The 5/3 bilayer that we
choose for this investigation contains two edge planes, two middle planes, and one
center plane in the AF1 layer and two edge planes and one center plane in the AF2
layer [see Fig.~\ref{fig01}]. This provides a platform to estimate the range of the
interfacial effect, i.e., how far AF1 layer affects the AF2 layer and vice versa. In
Figs.~\ref{fig03}(a) and (b), we show the moment profiles of the constituent planes of
the AF1 and AF2 layers, respectively at $T = 0.02$ by varying the strength of Coulomb
repulsion $U_{2}$ of the AF2 layer. In the center plane of the AF1 layer, the magnetic
moment stays almost constant and near the bulk limit for $U_1 = 8$ [see Fig.~\ref{fig03}(a)].
Additionally, the magnetic moment in the middle planes follows the center plane and
does not fluctuate significantly. But, the presence of AF2 layer modifies the magnetic
moment in the AF1 layer's edge planes. For $U_{2} < 8$ ($U_{2} > 8$), the average
moment of the AF1 layer's edge planes is smaller (greater) than that of the layer's
center plane. For visual clarity, we have displayed
$dM_{AF1}$ ($= M_{AF1}^{center} - M_{AF1}^{edge}$) in the inset of Fig.~\ref{fig03}(a)
to illustrate the crossover of $dM_{AF1}$ from positive to negative at $U_{2}$ = $U_{1}$ = 8.
On the other hand, the AF2 layer's center plane and edge plane moments grow monotonically
as $U_{2}$ increases and saturate at higher $U_{2}$ values (see Fig.~\ref{fig03}(b))
similar to the bulk calculations~\cite{Mukherjee,Mondaini}. Unlike the AF1 layer, the
average moment in the edge plane is bigger (smaller) than the center plane for
$U_{2} < 8$ ($U_{2} > 8$).
However, the difference is very small for $U_{2} > 8$. The negative to positive
crossover of $dM_{AF2}$ ($= M_{AF2}^{center} - M_{AF2}^{edge}$) is shown
in the inset of Fig.~\ref{fig03}(b).

It is highly likely that the coupling between antiferromagnetic layers at the
interface in AF1/AF2 bilayers alters the magnetic moment profiles of edge planes.
To investigate this correspondence, we evaluate the effective hopping $t_{eff}$ for
the various planes of the AF1 and AF2 layers, as illustrated in Figs.~\ref{fig03}(c)
and (d), respectively, for 5/3 bilayer at $T = 0.02$. The $t_{eff}$ values remain
constant in the center and middle planes of the AF1 layer as we increase $U_{2}$
values and the obtained values are closer to the bulk limit for $U_{1} = U = 8$
[see Fig.~\ref{fig03}(c)]. This indicates that the moment profile of central planes
remains more or less unaffected.

The $t_{eff}$ in both edge and center planes of the AF2 layer decreases monotonically
with increasing $U_{2}$ values as shown in Fig.~\ref{fig03}(d). This pattern indicates
that moments are becoming increasingly localized with increasing $U_{2}$.
This decrease in $t_{eff}$ value with increasing $U_{2}$ has a significant impact on the
edge plane of the AF1 layer as shown in Fig.~\ref{fig03}(c).
Unlike the AF1 layer, the $t_{eff}$ of the edge plane of the AF2 layer is smaller (bigger)
than that of the center plane for $U_{2} < 8$ ($U_{2} > 8$).
As a result, for $U_{2} < 8$ ($U_{2} > 8$), the moments in the edge plane of AF2 layer
become more (less) localized than the center plane when the AF2 layer makes contact with
the AF1 layer. Overall, the $t_{eff}$ and $M$ calculations shown in Fig.~\ref{fig03},
corroborate each other very nicely.

\begin{figure}[!t]
\centerline{
\includegraphics[width=8.5cm,height=6.30cm,clip=true]{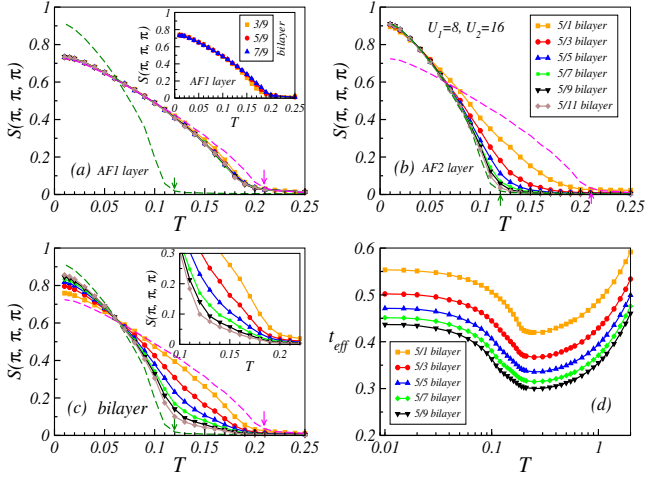}}
\caption{
S$(\pi, \pi, \pi)$ vs $T$ across various layers: (a) the AF1 layer, (b)the AF2 layer,
and (c) the whole bilayer. Thickness of AF1 layer is set at $w_{1} = 5$, whereas the AF2 layer's
thickness varies from $w_{2} = 1$ to $11$. The magneta and green dashed lines are used to plot the
S$(\pi, \pi, \pi)$ vs $T$ of bulk systems corresponding to $U_1 = 8$ and $U_2 = 16$, respectively.
The magneta and green arrows represent the $T_{N}$s of the bulk systems. The $T_N$ of the AF1
layers remains same for all the bilayers. The inset of (a) displays the temperature dependence
of S$(\pi, \pi, \pi)$ of the AF1 layer in 3/9, 5/9, and 7/9 bilayers. Inset also shows that the
$T_N$ of the AF1 layers remains intact when we fix $w_{2} = 9$ and vary $w_{1}$. However, the
$T_{N}$ of the AF2 layer decreases as its thickness increases. Interestingly, the $T_{N}$ of
the entire bilayer stays more or less constant. The inset of (c) displays an enlarged view of
the structure factors of the whole bilayers close to the transition temperature. In (a)-(c),
legends are the same. (d) The temperature evolution of $t_{eff}$ for different bilayers is
displayed. The dip points in $t_{eff}$, which represent the antiferromagnetic transition, are
the same for each bilayer. As the thickness of the AF2 layer varies, this reiterates
that the $T_{N}$ of the whole bilayer remains unchanged. Throughout all calculations, $U_{1}$
(associated with the AF1 layer) and $U_{2}$ (linked to the AF2 layer) are set at $U_{1} = 8$
and $U_{2} = 16$.
} 
\label{fig04}
\end{figure}

\section{Magnetotransport properties of AF1(high-$T_{N}$)/AF2(low-$T_{N}$) bilayers
  [$U_{1} = 8$ and $U_{2} = 16$]}

Now, we study the AF1/AF2 bilayers in details by assigning $U_{2} = 2\times U_{1}$,
for which $T_{N}$ of both AF1 and AF2 are clearly distinct. For $U_1 = 8$, the $T_{N}$
of corresponding bulk system is $0.21$, with a magnetic moment $M$ of roughly 0.86. 
For $U_2 = 16$, the corresponding bulk system has a lower $T_{N}$ (= $0.12$) and
a magnetic moment $M$ of approximately $M \sim 0.95$. In our calculations, we classify
AF1 (with $U_1 = 8$) and AF2 (with $U_2 = 16$) layers as high-$T_{N}$ and low-$T_{N}$
antiferromagnetic layers, respectively. The parameter selection reflects the qualitative
characteristics of antiferromagnetic layers in well-studied NiO/CoO bilayers,
including higher $T_{N}$ for NiO and higher magnetic moment for CoO~\cite{Jena1,Li,JZhu,Borchers1,Takano}.
In addition, these $U$ values comply with those calculated in density functional
theory calculations for NiO~\cite{Yu,Hermawan,Dudarev,Rohrbach,Zhang}and CoO~\cite{Kostov,Saha}.
So, we choose these two $U$ values to demonstrate the enhancement of Neel temperature of
a low-$T_{N}$ antiferromagnet in contact with a high-$T_{N}$ antiferromagnet, as observed
in NiO/CoO bilayers~\cite{Jena1,Li,JZhu,Borchers1,Takano}. It is important to note here
that our calculations for mimicking NiO or CoO are qualitative in nature.

The magnetic properties of AF1/AF2 bilayers are analyzed by calculating the antiferromagnetic
structure factors S$(\pi, \pi, \pi)$ for the layers and the planes, as well as the total
bilayer. Temperature-dependent structure factors S$(\pi, \pi, \pi)$ for bilayers and
individual layers by changing the thickness of the AF2 layer ($w_{2}$) while keeping the
thickness of the AF1 layer constant at $w_{1} = 5$ are shown in Fig.~\ref{fig04}. The $T_{N}$
of the AF1 layer stays stable at $T \sim 0.2$ in all bilayer systems investigated in this
work as shown in Fig.~\ref{fig04}(a). The $T_N$ of the AF1 layer also remains constant
when the thickness of the AF1 layer ($w_{1}$) is modified, but the thickness of the AF2
layer is fixed, as illustrated in the inset. On the other hand, $T_{N}$ of the AF2 layer
decreases as the thickness $w_{2}$ increases, approaching the bulk limit at $w_{2} = 11$
[see Fig.~\ref{fig04}(b)]. Interestingly, as the thickness of the AF2 layer increases,
the bilayer's $T_{N}$ remains relatively constant due to the influence of the AF1 layer
[see Fig.~\ref{fig04}(c)]. The zoomed version around $T_N $ is plotted in the
inset. We also ascertain the $T_N$ by calculating the effective hopping parameter
$t_{eff}$ of the bilayers. The localization of moments causes a decrease in $t_{eff}$
as the temperature decreases. It then starts increasing at $T_{N}$ due to delocalization
of moments supported by virtual hopping, which establishes the onset of the
antiferromagnetic ordering. A dip in $t_{eff}$ around $T = 0.2$ as shown for different
bilayers in Fig.~\ref{fig04}(d) helps us to double check the $T_{N}$ values.

\begin{figure}[!t]
\centerline{
\includegraphics[width=8.5cm,height=6.30cm,clip=true]{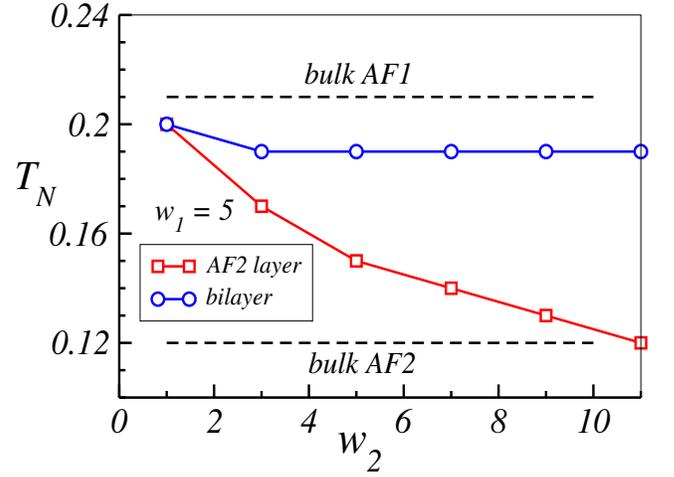}}
\caption{
The Neel temperature $T_{N}$ of the AF2 layer and the whole bilayer is displayed for various
AF2 layer thicknesses. While the $T_{N}$ of the AF2 layer decreases as its thickness increases,
the $T_{N}$ of the bilayer remains more or less constant. The black dashed lines represent
the $T_{N}$s of the bulk systems corresponding to the AF1 ($U_{1} = 8$) and AF2 ($U_{2} = 16$)
layers as indicated in th figure.} 
\label{fig05}
\end{figure}

Fig.~\ref{fig05} summarizes the systematics of the $T_{N}$. The black dashed lines
represent the $T_N$s of the bulk systems corresponding to the AF1 ($U_1$ = 8) and
AF2 ($U_2$ = 16) layers. Overall, our calculations show that the $T_{N}$ of
the AF2 layer decreases with increasing the thicknesses of the AF2 layer in the AF1/AF2
bilayer system. So, the $T_{N}$ of the low-$T_{N}$ AF2 layer increases significantly
at smaller thicknesses, whereas at larger thicknesses, it approaches the value of
the bulk system.

\begin{figure}[!t]
\centerline{
\includegraphics[width=8.5cm,height=6.30cm,clip=true]{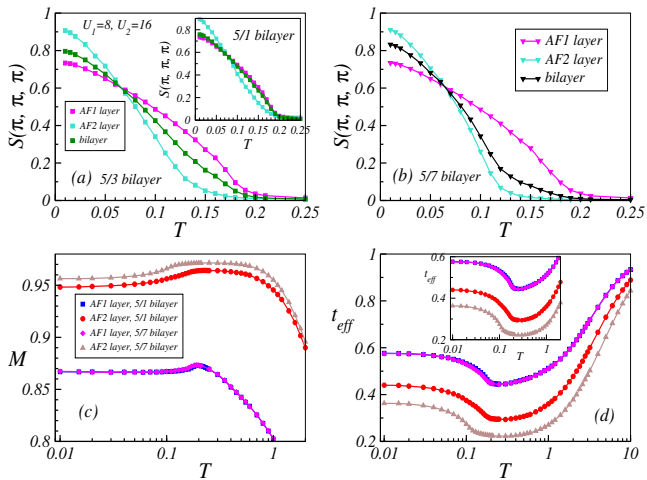}}
\caption{
\label{fig06}
Layer-resolved magnetic properties: Temperature variation of the structure factor
$S(\pi, \pi, \pi)$ of AF1 and AF2 layers along with the total bilayer are shown for
(a) the 5/3 bilayer and (b) the 5/7 bilayer. The $T_N$ of the AF2 layer in 5/3 bilayer
is enhanced whereas the $T_N$ of the AF2 layer in 5/7 bilayer approaches its bulk limit.
Inset of (a) demonstrates that the $S(\pi, \pi, \pi)$ of the AF1 and AF2 layers as well
as the total 5/1 bilayer exhibit same $T_N$.
(c) Temperature evolution of the local moment $M$ of the AF1 and AF2 layers is plotted
for the 5/1 and 5/7 bilayers. A little decrease in moment is observed for the AF2 layer
in the 5/7 bilayer at a lower temperature (see range $T = 0.1-0.2$) compared to the
temperature for the AF2 layer in the 5/1 bilayer. This indicates that the AF2 layer's
$T_N$ in the 5/7 layer is less than that of the 5/1 bilayer. For more details please
see the texts.
(d)For the 5/1 and 5/7 bilayers, the temperature evolution of $t_{eff}$ of the
constituent AF1 and AF2 layers is displayed. The low temperature upturn in $t_{eff}$
of the AF2 layer, which is associated with Neel temperature, is situated lower temperature
in the 5/7 bilayer than in the 5/1 bilayer. The inset shows a zoomed version of the same 
In (c) and (d), legends are same. All the calculations are performed for $U_{1} = 8$
and $U_{2} = 16$.
} 
\end{figure}

Additionally, for better visualization, we plot the S$(\pi, \pi, \pi)$ vs $T$ for the
individual AF1 and AF2 layers alongside their associated bilayers in
Fig.~\ref{fig06}(a) and Fig.~\ref{fig06}(b) for 5/3 and 5/7 bilayer, respectively. In the
inset of Fig.~\ref{fig06}(a), we illustrate the $T_{N}$ of the AF1 and AF2 layers for 5/1 bilayer
system, where the $T_{N}$ of the AF2 layer closely matches that
of the AF1 layer. In 5/3 bilayer, the $T_N$ of AF1 and AF2 layers differ significantly, while
the $T_N$ of the AF2 layer stays higher than that of the bulk counterpart. Hence, the proximity
of the high-$T_{N}$ antiferromagnet in the bilayers enhances its $T_{N}$,
particularly for thinner AF2 layers, similar with experiments~\cite{Jena1,Li,JZhu,Borchers1,Takano}.
For 5/7 bilayers, the bifurcation between $T_N$ of AF1 and AF2 layers is apparent more clearly.
The $T_{N}$ of the AF2 layer in 5/7 bilayer is close to the bulk value $T_{N}\sim 0.12$ and
for the bilayer that has thicker AF2 layer (e.g. $w_{2} = 11$), the $T_{N}$ of the AF2 layer
matches to its bulk limit [see Fig.~\ref{fig04}(b)]. Therefore, it is obvious that the proximity
effect of the AF1 layer has a considerable impact on the $T_N$ of the AF2 layer up to $w_{2} = 9$.

Furthermore, we show the moment profiles of the individual AF1 and AF2 layers for the
5/1 and 5/7 bilayers in Fig.~\ref{fig06}(c) in order to obtain additional evidence of
the thickness dependency of the $T_N$. It is well known that the bulk system
exhibits a tiny peak around transition temperatures as a result of moment delocalization
aided by virtual hopping assisted by the antiferromagnetic correlations~\cite{Chakraborty1}.
The magnetic moment profile of the AF1 layers also shows a peak around $T$ = 0.2 that
is similar to bulk systems. The AF1 layer's peak location in the 5/1 bilayer and the 5/7
bilayer overlap with each other. As previously presented in Fig.~\ref{fig04}(a), this
also suggests that the resulting $T_N$ of the AF1 layer in 5/1 and 5/7 bilayers are
equal to each other. While the AF2 layer's peak structure is not very apparent, the
antiferromagnetic transition is indicated by a modest drop in moment between
$T = 0.2$ and $0.1$. As we move from the 5/1 to the 5/7 layer, the temperature at
which this drop is observed decreases, indicating that the $T_{N}$ of the AF2 layer
in the 5/7 bilayer is lower than that in the 5/1 bilayer.

In addition, we evaluate the effective hopping parameter $t_{eff}$ of AF1 and AF2
layers for 5/1 and 5/7 bilayers to ascertain the antiferromagnetic transitions.
First, as previously stated, the temperature at which the AF1 layers' dip in
$t_{eff}$ appears corresponds to their magnetic transition and comes as equal to
$T_N$. Secondly, the dip for AF1 and AF2 layers occurs at the same temperature
for 5/1 bilayers as shown in Fig.~\ref{fig06}(d), but differs for 5/7 layers. This
demonstrates that the $T_N$ of AF1 and AF2 layers (in 5/7 layers) are distinct
from one another. Consequently, the AF2 layer's $T_{N}$ in the 5/7 bilayer
is less than that in the 5/1 bilayer.

\begin{figure}[!t]
\centering
\includegraphics[scale=0.33]{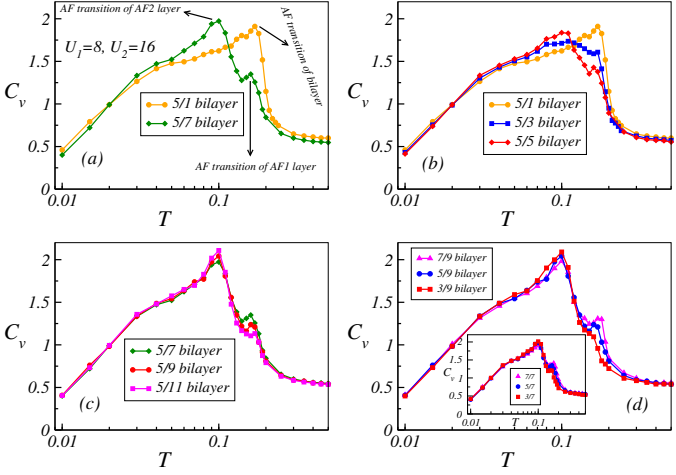}
\caption{
\label{fig07}
a) Temperature dependence of specific heat $C_{v}$ for the 5/1 and 5/7 bilayers. The specific
heat curve of the 5/1 (5/7) bilayer exhibits a single-peak (double-peak) structure.
A single peak in the 5/1 bilayer implies that the magnetic transition points for the AF1 and
AF2 layers are the same. One peak out of two peaks for the 5/7 bilayer (in particular, the peak
at relatively high temperature) is associated with the AF1 layer's antiferromagnetic transition,
while the other is associated with the AF2 layer's antiferromagnetic transition. Black arrows
point to these peaks.
(b) Temperature dependency of $C_{v}$ in 5/1, 5/3, and 5/5 bilayers is analyzed to identify
systematic variations as AF2 layer thickness varies.
(c) The specific heat $C_{v}$ is plotted as a function of temperature for the 5/7, 5/9, and 5/11
bilayers. As the thickness $w_{2}$ of the AF2 layer increases, the peak associated with the
magnetic transition of the AF1 layer is suppressed.
In (d) specific heat variations of 3/9, 5/9, and 7/9 bilayers with temperature are compared.
Here, the thickness of the AF1 layer $w_{1}$ varies from $3$ to $9$, while the thickness of the
AF2 layer set at $w_{2} = 9$.  The inset in (d) shows the $C_{v}$ vs $T$ for the 3/7, 5/7, and
7/7 bilayers (i.e., with a fixed AF2 layer thickness of $w_{2} = 7$ and variable AF1 layer
thicknesses). The magnitude of the peak linked to the AF1 layer decreases as its thickness
$w_{1}$ decreases.
}
\end{figure}

We also calculate the specific heat $C_{v}$, as shown in Fig.~\ref{fig07}(a), in order
to understand the systematics of the transition temperatures of the bilayers with
change of the thicknesses of the AF2 layers. To get an idea about the transition
temperatures, we examine the specific heat up to $T = 0.5$. There is a noticeable
single-peak structure in the specific heat for 5/1 bilayer. This single-peak structure
suggests that the whole system experiences antiferromagnetic transition at a particular
temperature. This result also confirms that the $T_N$ of AF1 and AF2 layers are equal
to each other in 5/1 bilayer. Conversely, the 5/7 bilayer exhibits two-peak features
because to its thick AF2 layer. One of the two peaks (peak at relatively higher
temperature) is linked to the AF1 layer's antiferromagnetic ordering, while the
other is connected to the AF2 layer's antiferromagnetic transition. Therefore, for
higher (lower) thicknesses of the AF2 layers in the bilayers, the two-peak (one-peak)
character of the specific heat is compatible with the systematics of the structure
factors as shown in Fig.~\ref{fig06}.

In order to substantiate the two-peak structure of specific heat, we vary the thicknesses
of AF2 layer as shown in Fig.~\ref{fig07}(b). As we mentioned above the single peak around
$T = 0.2$ is associated with $T_N$ of both AF1 and AF2 layers for 5/1 bilayer. For 5/3
bilayer another broad peak emerges just below $T = 0.2$ that is related to the onset of
magnetization in the AF2 layer which differs from $T_N$ of the AF1 layer characterized
by the peak around $T = 0.2$. This second peak is more prominent for 5/5 bilayer. At the
same time, the peak height associated with the AF1 layer decreases with increasing the
thicknesses of the AF2 layers. This trend is also clearly visible as we move from 5/7 to
5/11 bilayer [see Fig.~\ref{fig07}(c)].

On the other hand, for a fixed  thickness of the AF2 layer, e.g., $w_{2} = 9$ [see
Fig.~\ref{fig07}(d)] or $w_{2} = 7$ [see inset of Fig.~\ref{fig07}(d)], the peak height
associated with the AF1 layer decreases as the thickness of the AF1 layer decreases from
$w_{1} = 7$ to $3$, while the peak height associated with the AF2 layer remains more
or less unaffected. In fact, the peak in the specific heat associated with the antiferromagnetic
ordering of the AF1 layer is not prominent for the 3/9 bilayer [see Fig.~\ref{fig07}(d)].
But,  the antiferromagnetic transition is inherent in the AF1 layer of 3/9 bilayer
like the AF1 layer in the 5/9 and 7/9 bilayers, as confirmed from the structure factor
calculations shown in the inset of Fig.~\ref{fig04}(a). Overall, for larger thicknesses
of the AF2 layer, the specific heat of the bilayers exhibits two-peak characteristics,
consistent with the experimental observations.

\begin{figure}[!t]
\centering
\includegraphics[scale=0.33]{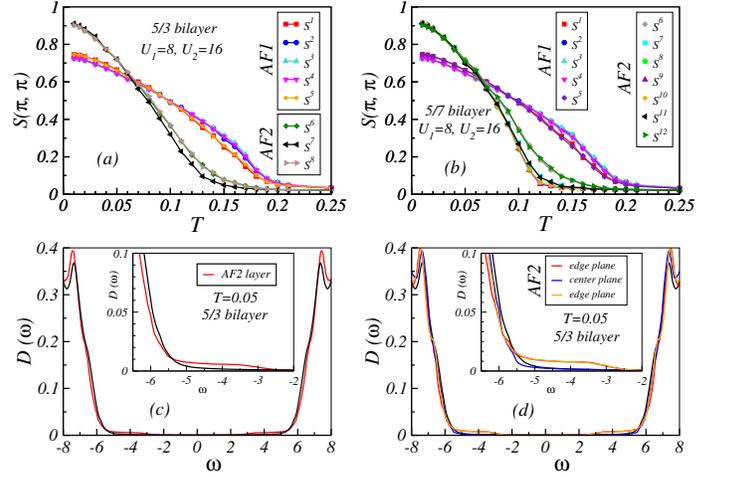}
\caption{
\label{fig08}
Temperature-dependent structure factor $S(\pi, \pi)$ for individual planes in (a) 5/3 and
(b) 5/7 bilayers. Antiferromagnetic transitions occur at $T \sim 0.2$, according to
$S(\pi, \pi)$ vs. $T$ plots for all the planes in the AF1 layer
(i.e. for $S^{1}$, $S^{2}$, $S^{3}$, $S^{4}$, and $S^{5}$ planes) for both the bilayers.
The $T_{N}$ of the edge planes ($S^{6}$ or $S^{8}$) of the AF2 layer in the 5/3 bilayer is
greater than that of the center plane ($S^{7}$). In 5/7 bilayer, as one moves deeper into
the AF2 layer from the interfaces (i.e., from $S^{6}$ to $S^{9}$), the $T_{N}$
decreases. The $S(\pi, \pi)$ of all the inner planes ($S^{8}$, $S^{9}$, and $S^{10}$)
exhibit antiferromagnetic transitions approximately at $T\sim 0.12$, which is equivalent
to the $T_{N}$ of the AF2 layer in its bulk limit.
(c) The density of states (DOS) of the AF2 layer of the 5/3 bilayer is displayed at
$T = 0.05$. The inset re-plots a zoomed-in version of the DOS close to the gap's boundary.
(d) At $T = 0.05$, the DOS of the AF2 layer's two edge planes and one center plane are
displayed separately. The edge planes shows finite although very small finite DOS at
the boundaries of the gap. Inset: The DOS close to a gap boundary is also re-plotted
in a zoomed-in version. For comparison, the DOS of the bulk system for $U_2 = 16$ is
also plotted at $T = 0.05$ by a black line in (c) and (d).
}
\end{figure}

We also explore the plane-resolved antiferromagnetic structure factors to better understand
how the AF2 layer's Neel temperature is enhanced. Figs.~\ref{fig08}(a) and ~\ref{fig08}(b)
show plots of $S(\pi, \pi)$ for all planes of the 5/3 and 5/7 bilayers, respectively. The
high-$T_{N}$ AF1 layer's constituent planes undergo antiferromagnetic transitions at
the same temperature, $T \sim 0.2$, for both the 5/3 and 5/7 bilayers. The edge planes of
low-$T_{N}$ AF2 layer in 5/3 bilayer display a higher Neel temperature than the center
plane. $S^{6}$ and $S^{8}$ represent the edge planes of the AF2 layer as indicated
in Fig.~\ref{fig08}(a). Nonetheless, the center plane ($S^{7}$ plane) has an enhanced
Neel temperature than the $T_N$ of AF2 layer at the bulk limit.

The inner planes of the AF2 layer in the 5/7 layer [$S^{8}$, $S^{9}$, and $S^{10}$ planes in
Fig.~\ref{fig08}(b)] exhibit antiferromagnetic transitions at the same temperature, i.e. at $0.12$,
which corresponds to the $T_{N}$ of the low-$T_{N}$ AF2 layer in the bulk limit. But, the $T_N$
of edge planes [see $S^{6}$ and $S^{12}$ planes in Fig~\ref{fig08}(b)] are truly enhanced, while
planes just next to the edge plane [see $S^{7}$ and $S^{11}$ plots in Fig.~\ref{fig08}(b)] are
marginally augmented as compared to inner planes. As the thickness of the AF2 layer increases,
the proximity effect is limited to its interfacial planes. It increases the Neel temperature
of the edge plane while leaving the inner individual planes unchanged. So, the $T_N$ decreases
as one advances deeper planes into the AF2 layer from the interface, eventually approaching the
bulk limit for the inner planes.

We calculate the density of states (DOS) to emphasize the fact that the edge planes are more
influenced. First, we plot the DOS of the AF2 layer for the 5/3 bilayer at $T = 0.05$ in
Fig.~\ref{fig08}(c).
Additionally, we plot the DOS of the corresponding bulk system for $U_2 = 16$. Like the bulk
system, the AF2 layer's DOS exhibits a pronounced Mott-gap. The main difference is that, for
the AF2 layer, a small finite DOS is visible at the gap's boundaries [see the inset of a zoomed
version]. To determine which planes of the AF2 layer are responsible for generating finite DOS on
the end of the gap, we examine the DOS of every single plane of the AF2 layer, as shown
in Fig.~\ref{fig08}(d). The DOS of the center plane roughly reciprocate with the DOS of the
corresponding bulk system. However, the AF2 layer's edge planes show finite DOS at the
gap's boundaries. This change in the DOS of the AF2 layer's edge planes results from the
coupling of the AF2 layer with the AF1 layer at the interfaces, which is driven by
delocalization of moments across the interfaces.

\begin{figure}[!t]
\centering
\includegraphics[scale=0.33]{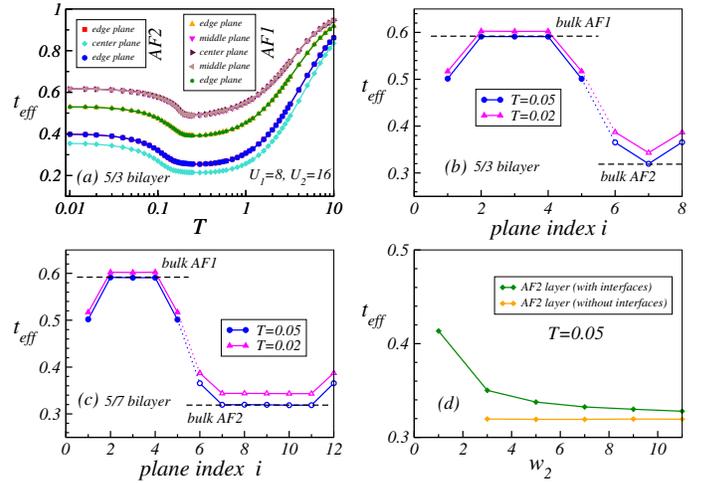}
\caption{
\label{fig09}
For the 5/3 bilayer, the temperature evolution of the $t_{eff}$ of the two edge planes,
two middle planes, and one center plane of the AF1 layer and the two edge planes and
one center plane of the AF2 layer is displayed. The AF1 (AF2) layer's edge planes have
a smaller (bigger) $t_{eff}$ than the center plane. $t_{eff}$ is plotted against the
plane index $i$ for (b) 5/3 bilayer and (c) 5/7 bilayer at $T = 0.05$ and
$0.02$. The planes of the AF1 (AF2) layer are represented by the solid (open) symbol.
The coloured dotted line shows the link between the interfacial planes. The black dashed
line represents the $t_{eff}$ of the bulk systems corresponding to AF1 ($U_1 = 8$) and
AF2 ($U_2 = 16$) layers at $T = 0.05$. This indicates that the inner plane $t_{eff}$
values of the AF1 and AF2 layers are equal to those of the respective bulk systems.
(d) The average value of the $t_{eff}$ in the AF2 layer (with or without interfacial
planes) is depicted at $T = 0.05$. As the AF2 layer thickness increases, the $t_{eff}$
value decreases. However, except for the interfacial planes, the $t_{eff}$ remains constant.
}
\end{figure}

In order to ascertain the extent to which the AF1 layer affects the AF2 layer regarding the
delocalization of moments across the interface, we calculate the $t_{eff}$ values of the
various planes for the 5/3 and 5/7 bilayers. For the 5/3 bilayer, the temperature-dependent
$t_{eff}$ value [see Fig.~\ref{fig09}(a)] indicates that all the planes undergo a magnetic
transition as discussed in Fig.~\ref{fig09}(d). It also demonstrates that $t_{eff}$ in the
edge planes are bigger (smaller) than the center plane in both AF2 (AF1) layers.

We extract the $t_{eff}$ vs. plane index $i$ for the 5/3 bilayer at low temperatures
$T =  0.05$ and $0.02$ and plot those in Fig.~\ref{fig09}(b) to present a clear picture.
The same data are provided for the 5/7 bilayer as well in Fig.~\ref{fig09}(c). Because
of coupling with the strongly localized AF2 layer ($U_{2} = 16$) at the interfaces, it
is evident that at any given temperature, the $t_{eff}$ in the AF1 layer's edge planes
is less than the middle and center planes.  On the other hand, because of the coupling
with the relatively delocalized AF1 layer, the $t_{eff}$ in the AF2 layer's edge planes
is greater than the center plane. In other words, in the AF2 (AF1) layer, the moments
in the edge plane become more (less) delocalized than those in the center plane. The
enhancement of the $T_N$ of AF2 layer results from this delocalization-driven interaction
of the edge planes of the AF2 layer with the AF1 layer.

It's also interesting to note that while the average $t_{eff}$ in the AF2 layer
excluding the interfacial planes stays constant during the thickness variation of AF2 layer,
the average $t_{eff}$ of all the planes in the AF2 layer falls as its thickness increases
[see Fig.~\ref{fig09}(d)]. According to all of these results, the AF1 layer's
delocalization-induced influence on the AF2 layer decreases as the AF2 layer's thickness
increases and proximity effect mostly stays confined at the interfacial edge planes.

\begin{figure}[!t]
\centering
\includegraphics[scale=0.33]{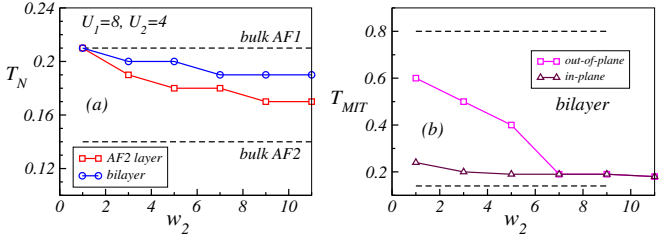}
\caption{
\label{fig10}
The $T_{N}$ of the AF2 layer and the whole bilayer are shown by varying the thicknesses of
AF2 layer. The $T_{N}$ of AF1 ($U_{1} = 8$) and AF2 ($U_{2} = 4$) layers in their bulk limit
are represented by the black dashed lines as indicated in the figure. The $T_{N}$ profile
clearly shows that the $T_{N}$ of the AF2 layer in AF1/AF2 bilayer is greater than that
of the corresponding bulk system.
(b) The $T_{MIT}$s of bilayers are plotted with varying AF2 layer thicknesses. The $T_{MIT}$
is determined by the temperature variations of the in-plane and out-of-plane resistivities.
The top (bottom) dashed line represents the $T_{MIT}$ of the bulk system corresponding to
the AF1 (AF2) layer. $U_{1}$ (associated with the AF1 layer) and $U_{2}$ (related with the
AF2 layer) are fixed at $U_{1} = 8$, $U_{2} = 4$ throughout the calculations.
}
\end{figure}

\section{Magnetotransport Properties of (High-$T_{N}$-AF)/(low-$T_{N}$-AF) bilayers
  [$U_{1} = 8$ and $U_{2} = 4$]} 

In previous section we explored the (high-$T_{N}$ AF1)/(low-$T_{N}$ AF2) bilayers by
assigning $U_{1} = 8$ to the AF1 layer and $U_{2} = 16$ to the AF2 layer, i.e., for the
$U_{2} = 2\times U_{1}$ scenario. For the $U_{2} = 0.5\times U_{1}$ case, we select
$U_{2} = 4$ with a bulk $T_{N}$ of $0.14$. This ensures that the AF2 layer acts as
a low-$T_{N}$ antiferromagnet. So, in this section we investigate the magnetic and
transport properties of another set of high-$T_{N}$/low-$T_{N}$ (AF1/AF2) bilayers
and compare them to prior section results as needed. Unlike the previous case, the
moments in the AF2 layer are more delocalized than in the AF1 layer. Thus, it will
be interesting to find out how the $T_N$ of the AF2 layer is impacted by the localized
moments in the AF1 layer. In our calculations, we modify the thickness of the AF2
layer while keeping the thickness of the AF1 layer fixed at $w_1 = 5$, just like we
studied in the previous section.

In Fig.~\ref{fig10}(a), we show how the $T_N$ of AF1/AF2 bilayer systems varies by
altering the thickness of the low-$T_{N}$ AF2 layer while keeping the thickness of the
high-$T_{N}$ AF1 layer constant at $w_{1} = 5$. The bilayer's $T_{N}$ decreases slightly
as the thickness of the AF2 layer increases. In the 5/1 and 5/3 bilayers
(i.e., for $w_{2}$ = 1 and 3), the low-$T_{N}$ AF2 layer's $T_{N}$ is significantly
enhanced and approaches that of the high-$T_{N}$ AF1 layer. Interestingly, the $T_{N}$
of the AF2 layer in the bilayer diminishes slowly as its thickness increases. This in
contrast with the rapid decrease in $T_N$ for the AF2 layer when $U_{2} = 16$ was used.
So, our calculations show that the value of $T_{N}$ of AF2 layer is higher than that of
the similar bulk system with $U_{2} = 4$.

\begin{figure}[!t]
\centerline{
\includegraphics[width=8.5cm,height=6.30cm,clip=true]{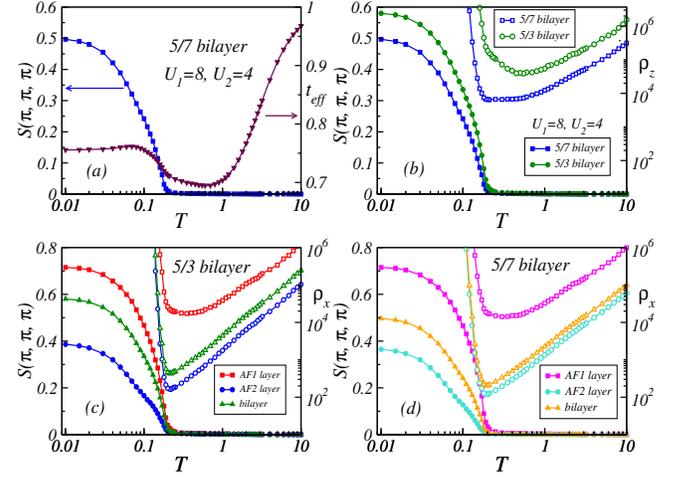}}
\caption{
(a) The temperature evolution of $t_{eff}$ and $S(\pi, \pi, \pi)$ are shown for 5/7 bilayer.
The $T_{N}$ is correlated with the $t_{eff}$ showing an upturn around $T = 0.2$. See text
for details.
(b) The temperature dependence of $S(\pi, \pi, \pi)$ and out-of-plane resistivity $\rho_{z}$
are plotted for the 5/7 and 5/3 bilayers. The 5/7 bilayer has a slightly smaller $T_{N}$ than
the 5/3 bilayer. While the 5/3 bilayer's $T_{MIT}$ is more than its $T_{N}$ value, the 5/7
bilayer's $T_{MIT}$ is equal to its $T_{N}$. Layer-resolved $S(\pi, \pi, \pi)$ and in-plane
resistivity $\rho_{x}$ for 5/3 and 5/7 bilayers are plotted in (c) and (d), respectively.
To ensure completeness, plots of $S(\pi, \pi, \pi)$ and $\rho_{x}$ for whole bilayers are
also shown.
} 
\label{fig11}
\end{figure}

Next, we show the metal-insulator transition temperature ($T_{MIT}$) of bilayer systems
in Fig.~\ref{fig10}(b). As the thickness of the AF2 layer increases, the $T_{MIT}$ obtained
from evaluating the resistivity in out-of-plane direction decreases. The $T_{MIT}$ ranges
from $0.8$ to $0.14$, representing the corresponding bulk systems with $U_{1} = 8$ and
$U_{2} = 4$, respectively. The range is illustrated by two dashed lines in Fig.~\ref{fig10}(b).
Beyond 5/7 bilayers, the $T_{MIT}$ equals $T_{N}$, but for other bilayers (5/1, 5/3, and 5/5),
$T_{MIT}$ exceeds $T_{N}$. In-plane resistivity calculations show that increasing the
thickness of the AF2 layer has little effect on $T_{MIT}$, except for the 5/1 bilayer with
a thin AF2 layer.

We present the resistivity data and antiferromagnetic structure factors S$(\pi, \pi, \pi)$
in Fig.~\ref{fig11} that were utilized to generate the summary shown in Fig.~\ref{fig10}.
We calculate the temperature dependent S$(\pi, \pi, \pi)$ and plot it with $t_{eff}$ of the
5/7 bilayer to ascertain the $T_{N}$ in Fig.~\ref{fig11}(a). The $T_N$ from S$(\pi, \pi, \pi)$
correlates with the $t_{eff}$, which shows an upturn around $T = 0.2$ as temperature decreases.
We also plot S$(\pi, \pi, \pi)$ for 5/3 and 5/7 bilayers in Fig.~\ref{fig11}(b) for comparison.
The $T_{N}$ of the 5/7 bilayer is slightly lower than the 5/3 bilayer but is much larger
than the corresponding bulk system for $U_{2} = 4$. The resistivity calculations in the
same figure reveal that the $T_{MIT}$ of the 5/7 bilayer matches with the $T_{N}$, whereas
in the 5/3 bilayer, the $T_{MIT}$ ($\sim 0.5$) is larger than the $T_{N}$. The fact that
the $T_{MIT}$ of 5/3 bilayer is different from the $T_{N}$ suggests that the AF1 layer plays
a crucial part in determining the transport in the bilayer when the AF2 layer is thin.
It is noteworthy to recall that, in the bulk case for $U_{2} = 4$, the $T_{MIT}$ and
$T_N$ are identical, while the $T_{MIT}$ of the bulk systems corresponding to $U_{1} = 8$
is larger than the $T_N$.

For each layer of the 5/3 bilayer, we show temperature dependence of S$(\pi, \pi, \pi)$ in
Fig.~\ref{fig11}(c). Additionally, for comparison, we plot the S$(\pi, \pi, \pi)$ for whole
bilayers. The AF2 layer's $T_{N}$ ($\sim 0.19$) is marginally less than the AF1 layer's
$T_{N}$ ($\sim 0.2$). Thus, for $U_{2} = 4$, the AF2 layer's $T_{N}$ is larger than its
corresponding bulk $T_{N}$, but it is still somewhat smaller than the $T_{N}$ of the AF1
layer. Fig.~\ref{fig11}(c) also displays the in-plane resistivities of the AF1 and AF2
layers as well as the entire 5/3 bilayer. The AF2 layer's $T_{MIT}$ is equal to the AF2
layer's $T_{N}$, while the AF1 layer's $T_{MIT}$ is greater than the AF1 layer's
$T_{N}$. The whole bilayer system exhibits a metal-insulator transition in its
in-plane resistivity at the same temperature as the $T_{N}$ and $T_{MIT}$ of AF2 layer.
Layer resolved magnetotransport properties of 5/7 bilayer are shown in Fig.~\ref{fig11}(d),
and the systematics of results are qualitatively similar to those of 5/3 layers.

\begin{figure}[!t]
\centering
\includegraphics[scale=0.33]{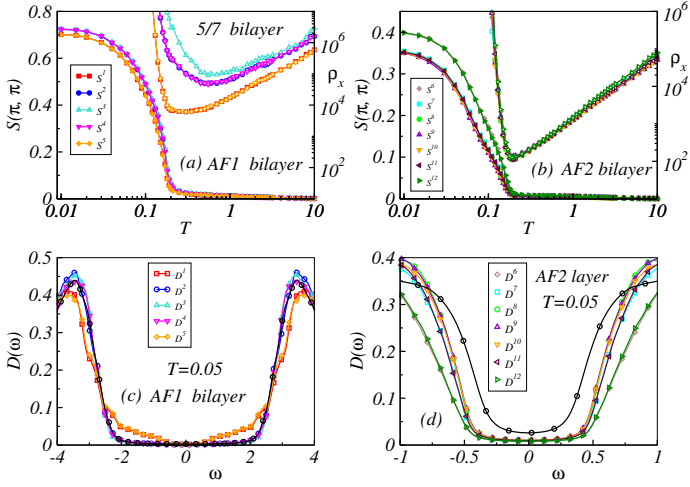}
\caption{
The temperature evolution of structure factor $S(\pi, \pi)$ and in-plane resistivity
$\rho_{x}$ for various planes in the AF1 and AF2 layers of 5/7 bilayer are shown in (a) and (b),
respectively. The structure factor (resistivity) curve is indicated by filled (open) symbols.
The individual planes of the AF1 layer ($S^{1}$, $S^{2}$, $S^{3}$, $S^{4}$, and $S^{5}$) exhibit
antiferromagnetic transitions at the same temperature, $T \sim 0.2$. The $T_{MIT}$ of the AF1
layer's edge planes is near to $T_{N}$, but for inner planes, it is bigger than $T_{N}$. In
the AF2 layer, the $T_{N}$ of the edge plane (from $S^{6}$ or $S^{12}$) is somewhat higher
than that of the inner planes (from $S^{7}$, $S^{8}$, $S^{9}$, $S^{10}$, $S^{11}$).
The resistivities of the AF2 layer's planes are very similar, and the $T_{MIT}$ matches
with the $T_{N}$. For the 5/7 bilayer, the density of states (DOS) of the constituent planes of
(c) the AF1 layer and (d) the AF2 layer are shown at $T = 0.05$. The DOS of the corresponding
bulk system is also plotted for $U_{1} = 8$ in (c) and $U_{2} = 4$ in (d) using a black line. While
the DOS in the inner planes ($D^{2}$, $D^{3}$, and $D^{4}$) are extremely close to the bulk limit,
the DOS of the AF1 layer's edge plane ($D^{1}$ or $D^{5}$) is modified. The AF2 layer's planes
exhibit a wider gap around the Fermi level ($\epsilon_{F} = \omega = 0$) compared to the bulk
system. The edge planes (from $D^{6}$ and $D^{12}$) have a somewhat larger gap than the inner
planes ($D^{7}$, $D^{8}$, $D^{9}$, $D^{10}$, and $D^{11}$).
}
\label{fig12}
\end{figure}

\begin{figure}[!t]
\centering
\includegraphics[scale=0.31]{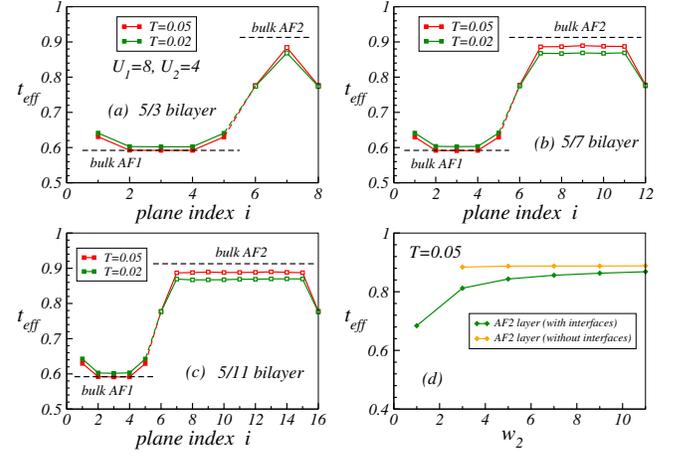}
\caption{
\label{fig13}
For the 5/3, 5/7, and 5/11 bilayers the $t_{eff}$ vs. plane index $i$ at $T = 0.05$ and $0.02$
are plotted in (a), (b), and (c), respectively.  The planes of the AF1 (AF2) layer are marked
by solid (open) symbols. The colored dashed line indicates the link between the interfacial
planes. The black dashed line represents the $t_{eff}$ of the bulk systems corresponding to
$U_1 = 8$ and $U_2 = 4$ at $T = 0.05$. The $t_{eff}$ of inner (edge) planes in the AF1 layer
is equal to (greater than) the bulk limit. This indicates that the AF1 layer's edge planes
are mostly affected in bilayer system. The AF1 layer affects all the planes of the AF2 layer,
resulting in smaller $t_{eff}$ values than the bulk value. However, the AF2 layer's edge planes
have smaller $t_{eff}$ values than the inner planes and are more affected due to interfacial
coupling.
(d) At $T = 0.05$, the average value of $t_{eff}$ is plotted in the AF2 layer and AF2 layer
without interfacial planes. The $t_{eff}$ in the AF2 layer increases with thickness and reaches
saturation at $w_{2} = 9$. However, except for the interfacial planes, the $t_{eff}$ of AF2
layer remains constant.}
\end{figure}

To get more insight of the proximity effect on the AF2 layer, we examine the plane-resolved
magnetotransport properties of the bilayer systems. We show the temperature dependence of
$S(\pi, \pi)$ and in-plane resistivity for each plane of the AF1 layer of the 5/7 bilayer
in Fig.~\ref{fig12}(a). The constituent planes of the high-$T_{N}$ AF1 layer undergo
antiferromagnetic transitions at roughly the same temperature, $T \sim 0.2$. However,
compared to other planes, the edge plane's $T_{MIT}$ is much closer to the $T_N$. This
indicates that the AF1 layer's edge plane is more impacted by the contact with the AF2 layer.
Meanwhile, the low-$T_{N}$ AF2 layer's edge planes show a marginally higher Neel
temperature than the center planes. $S^{6}$ and $S^{12}$ represent the edge planes of
the AF2 layer as indicated in Fig.~\ref{fig12}(b). The Neel temperatures of all inner
planes exceed the $T_N$ of the AF2 layer at the bulk limit. However, there are notable
similarities in the temperature dependence of the in-plane resistivity of the different
AF2 layer planes, and $T_{MIT}$ agrees well with the $T_N$ of the individual AF2 layer
planes.

We also plot the plane resolved density of states (DOS) for the AF1 and AF2 layers of
the 5/7 bilayer at $T = 0.05$ in Fig.~\ref{fig12}(c) and (d), respectively. For comparison,
we include the DOS of the bulk systems corresponding to AF1 ($U_{1} = 8$) and AF2 ($U_{2} = 4$)
layers. The DOS of the interfacial plane away from the Fermi level ($\epsilon_{F} = \omega = 0$)
is somewhat modified as shown in Fig.~\ref{fig12}(c). But, the gap around the Fermi
level for all the planes is essentially unchanged as compared to the bulk system, according
to the plane-resolved DOS of the high-$T_{N}$ AF1 layer. However, the DOS of all planes in
the AF2 layer shows a larger gap around the Fermi level than the bulk system as
illustrated in Fig.~\ref{fig12}(d). When the AF2 layer comes into contact with the AF1
layer with a large gap, the proximity effect increases the gap size of the AF2 layer. As
expected, proximity to the AF1 layer affects the edge plane of the AF2 layer more than the
inner planes.

Next, we plot the $t_{eff}$ vs. plane index $i$ for the 5/3 bilayer at low temperatures
$T = 0.05, 0.02$ as shown in Fig.~\ref{fig13}(a). The AF1 layer's edge plane has a
bigger $t_{eff}$ than its central (middle and center) planes, suggesting that the moments
there become more delocalized upon coming into contact with the AF2 layer. In fact, the AF1
layer's central planes are mostly unaffected, and $t_{eff}$ is near the comparable bulk value
for $U_1 = 8$. Conversely, because of the coupling with the more localized AF1 layer, the
$t_{eff}$ in the edge plane of the AF2 layer is smaller than that of the center plane.
In the center plane of the AF2 layer, the $t_{eff}$ is also smaller than the corresponding
bulk value of $U_2 = 4$.

Additionally, we present the $t_{eff}$ vs. plane index $i$ for the 5/7 and 5/11
bilayers at $T = 0.05, 0.02$ in Fig.~\ref{fig13}(b) and Fig.~\ref{fig13}(c)
respectively. These plots make it clear that all the planes inside the AF2 layer
are affected even for larger thicknesses of AF2 layer. However, the moments in
edge plane is more prone to localization than the inner planes. The average $t_{eff}$ in
the AF2 layer, excluding interfacial planes, remains constant as the inner planes are
equally affected due to the proximity effect of the AF1 layer, as shown in Fig~\ref{fig13}(d).
But, the average $t_{eff}$ of the total AF2 layer increases with $w_2$ (thickness of AF2 layer)
and saturates beyond $w_2 \sim 9$. These results comprehensively show that the all
the planes of AF2 layers are affected although edge plane is more influenced than
inner planes. Because of this, the $T_N$ of the AF2 layer (with $U_2 = 4$) in the
bilayer gradually decreases as its thickness increases, in contrast to the results
obtained when $U_2 = 16$ was assigned.

\section{Conclusions}

In summary, we investigated the magnetotransport properties of the
AF1(low-$T_{N}$)/AF2(high-$T_{N}$) bilayers using a one-band Hubbard model
at half-filling using semi-classical Monte Carlo approach. In our model Hamiltonian
calculations, we set Coulomb repulsion $U_{1}$ = 8 to simulate the high-$T_{N}$ AF1
layer, and $U_{2}$ = 2 $\times$ $U_1$ to simulate the low-$T_{N}$ AF2 layer. We choose
these parameters to mimic NiO/CoO like bilayers. Our calculations indicate that when
the thickness of the low-$T_N$ antiferromagnet is small, the proximity effect
significantly increases its $T_N$, resulting in a single magnetic transition temperature
for the bilayer system. A single peak in specific heat corresponds to the bilayer's
single-shot antiferromagnetic transition. As the thickness of the AF2 layer increases,
its $T_{N}$ decreases and approaches the bulk limit, indicating separation from the $T_N$
of the AF1 layer. The two-peak structure in the specific heat for thicker AF2 layers
supports these findings. In particular, we demonstrate that the increase in $T_N$ of
the AF2 layer nevertheless remains an interfacial effect. Overall, our findings
qualitatively agree with experimental results and provide insights into the phenomenon
of increasing the Neel temperature of low-$T_{N}$ antiferromagnets in bilayer systems.

We additionally carried out the analysis for $U_{1}$ = 8 and $U_{2} = 0.5\times U_{1}$
for completeness. Here, the low-$T_{N}$ antiferromagnet's $T_{N}$ is enhanced even for
thicker AF2 layers, unlike when $U_{2} = 2\times U_{1}$. Therefore, in the
$U_{2} = 0.5\times U_{1}$ instance, the proximity effect penetrates to the inner planes
due to a comparatively higher delocalization of moments in the AF2 layer. Density of states
calculations also show that proximity to the AF1 layer significantly impacts both
the interfacial and inner planes of the AF2 layer in this scenario. It would be
interesting to conduct experiments on these types of bilayers.

\begin{center}
\textbf{ACKNOWLEDGMENT}
\end{center}

We acknowledge use of the Meghnad2019 computer cluster at SINP.


\end{document}